\documentclass[11pt]{article}
\usepackage{blois2002,epsfig,pennames,eucal,array,amssymb}
\def\Journal#1#2#3#4{{#1} {\bf #2}, #3 (#4)}

\def\NIMA{{\em Nucl. Instrum. Methods} A}
\def\NPB{{\em Nucl. Phys.} B}

\def\PRD{{\em Phys. Rev.} D}

\def\ASJ{\em Astrophys. J.}

\def\be{\begin{equation}}
\def\ee{\end{equation}}
\def\bea{\begin{eqnarray}}
\def\eea{\end{eqnarray}}

\begin{document}
\vspace*{4cm}
\title{MEASUREMENT OF THE $\Pap/\Pp$ RATIO IN COSMIC RAYS USING THE MOON SHADOW
WITH THE L3+C DETECTOR}
\author{J.F. PARRIAUD\\
on behalf of the L3 collaboration}
\address{Institut de Physique Nucl\'eaire de Lyon, 4 rue Enrico Fermi,\\
69622 Villeurbanne, France}
\maketitle\abstracts{The observation of the Moon shadow effect with the L3+C 
detector is reported. The expected offset and elongation of the Moon shadow due to
the geomagnetic field are clearly observed. A determination of the angular 
resolution and pointing error has been performed through this 
observation. A preliminary upper limit of the $\Pap/\Pp$ ratio in cosmic rays is
given in the TeV region.}
\section{Introduction}\label{sec:intro}
In 1957, Clark~\cite{clark} postulated the effect of the Moon or the Sun on 
cosmic rays: as these bodies pass overhead during a transit, they block
particles, resulting in shadows in cosmic rays visible by detectors on Earth.
The observation of this shadow may be used to check the angular resolution of the
apparatus and to evaluate pointing errors.
Another use has been proposed in 1990~\cite{urban}: positively charged particles
are deflected by the Earth's magnetic field towards the East. Consequently 
their absorption by the Moon gives rise to a shadow on the West side of the 
Moon. Negatively charged primaries (antimatter) are deviated to the opposite 
side. A shadow on the East side of the Moon would therefore be observed in 
presence of antimatter in cosmic rays.\\
Up to now, no antinuclei heavier than antiprotons have been detected. Antiproton
flux measurements have been performed by balloon-borne experiments up to 40 GeV
(Figure 1). They are consistent with the antiproton production
expected from proton interaction in the interstellar medium:
$\Pp\,+\,N_{z}\,\rightarrow\,\Pap\,+\,X$. However one can notice that the
$\Pap/\Pp$ ratio measurements are still compatible with an increase expected
from an extragalactic component or heavy particle decays in the
galaxy~\cite{ullio}. 
Measurements at higher energies are thus welcome to set constraints on the 
existence of antimatter in the universe.\\
A description of the L3+C experiment~\cite{l3cnim} is given in section \ref{sec:l3c}.
In section \ref{sec:moon}, relevant aspects of the Moon shadow study are
detailed. Finally, in section \ref{sec:anal}, the procedure followed to analyse
the data is explained, a preliminary measurement of the $\Pap/\Pp$ ratio in 
the TeV region is performed and a 90\% C.L. upper limit is inferred.\\
\begin{center}
\hspace*{-1cm}\begin{tabular}[!ht]{c p{0.4cm} c}
  \hfill\includegraphics[width=7 cm,bb= 15 150 566 650]{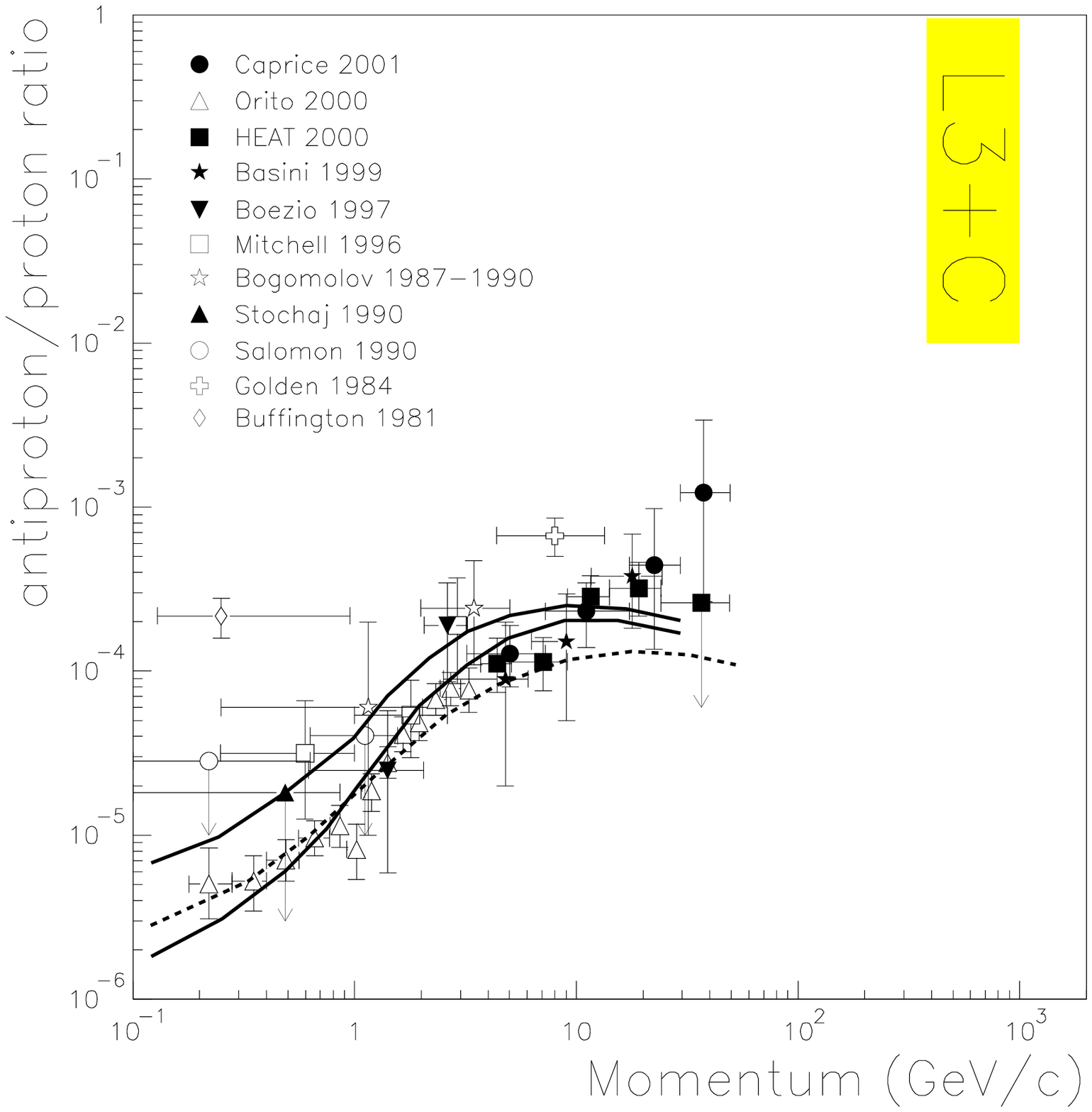}\hfill  &
  &
  \hfill\includegraphics[width=7 cm]{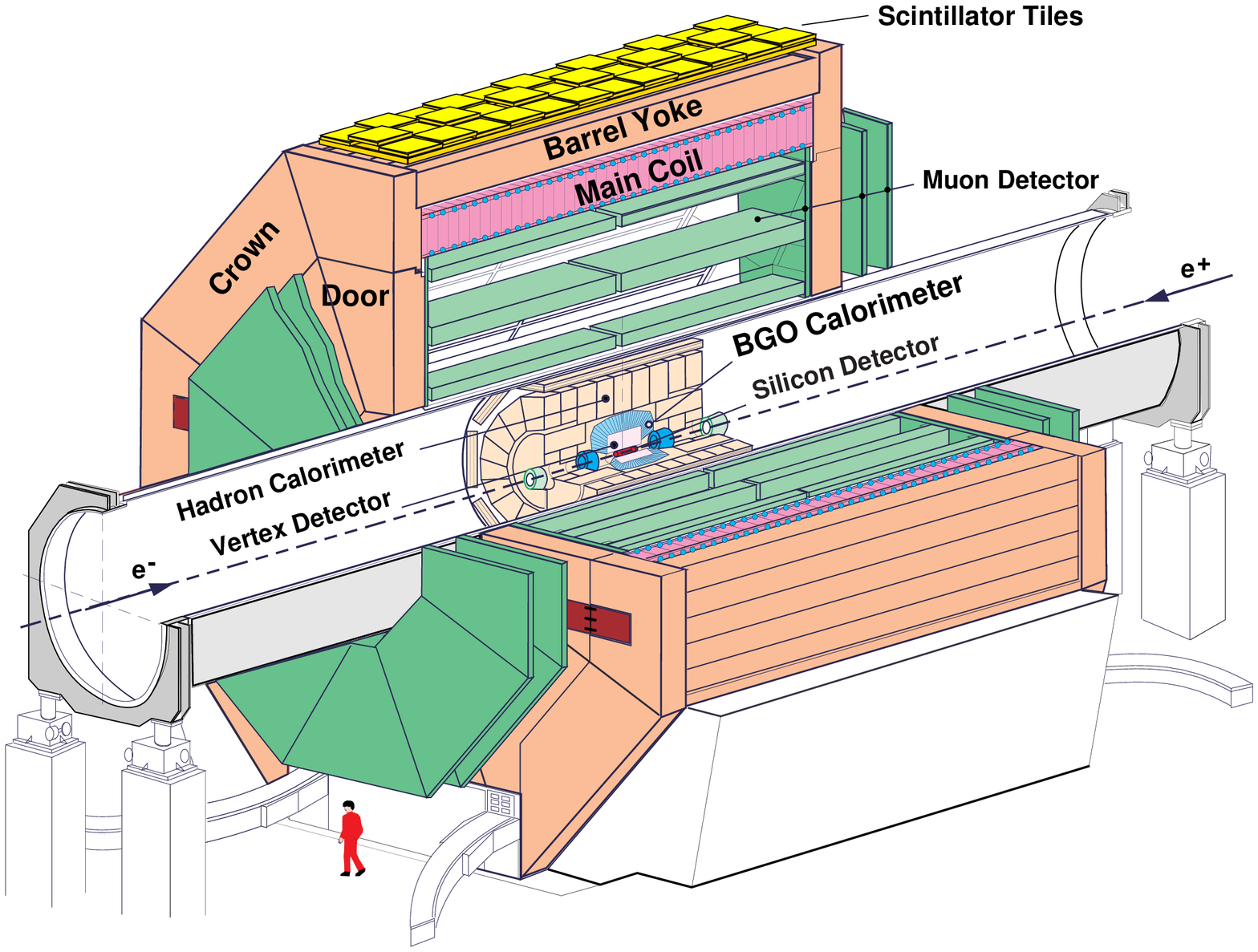}\hfill \\
\end{tabular}
\begin{tabular}[!ht]{p{7.5cm}p{0.2cm}p{6.8cm}p{1cm}}
  Figure 1 : \footnotesize{$\Pap/\Pp$ ratio versus momentum, adapted from
reference 5. The lines are calculations of interstellar antiprotons
assuming a pure secondary production during the propagation of cosmic rays in
the Galaxy. The L3+C experiment is sensitive to a $\Pap/\Pp$ ratio at a few
$\%$ level, for a primary proton energy between 0.5 TeV and 1 TeV.} &
  &
  Figure 2 : \footnotesize{The L3 detector. The solenoid, the muon chambers and
the scintillator tiles are the main ingredients of the L3+C detector. The
sub-detectors located inside the support tube are not used.} & \\
\end{tabular}
\end{center}
\addtocounter{figure}{2}
\section{The L3+C experiment}\label{sec:l3c}
L3 (Figure 2) was one of the four particle detectors installed on the Large 
Electron Positron collider (LEP) site, located underneath the French-Swiss 
border at CERN. It was designed to make a very precise 
measurement of leptons and photons produced in $\Pep\Pem$ collisions. In 
the late 1990s, it was updated to detect cosmic muons originating from the 
decay of charged pions produced when a high energy cosmic ray collides with an 
air nucleus.\\
One of the assets of the L3 detector was its muon spectrometer. It consisted of 
two octogonally shaped rings, each with eight "octants", located in the 1000 
$\rm m^{3}$ magnetic field of 0.5 T created by the solenoid. Each octant 
contained precision drift chambers organised in 3 layers to 
measure the projection of the muon trajectory onto the plane orthogonal to the
magnetic field, and 
layers of drift cells to measure the projection along the magnetic field 
direction. The momentum resolution achieved by this setup was 2.5\% at 45 GeV. 
Other parts of the L3 detector were not used by the L3+C experiment.
The reference time $t_{0}$, when a muon enters the detector, was measured by 
202 $\rm m^{2}$ of additional scintillators covering the upper part of the detector with a 1.5 
ns resolution. In order to register events independently of L3 running, new 
trigger and data acquisition system were made. Since the L3 detector was located 
under 30 m of molasse, the muon momentum threshold was about 15 GeV 
and it provided shielding from the electromagnetic and hadronic components of 
the air showers. In addition to this underground muon detector, an
Air Shower (EAS) array of 50 scintillators has been installed at the surface.\\
The original purpose of L3+C is the measurement of the muon momentum 
spectrum from 20 to 2000 GeV, which is related to the neutrino 
flux. Other topics are under study, among which the search for point sources 
and gamma-ray bursts, where the knowledge of the angular resolution and the 
pointing error is of primary importance. The study of the Moon 
shadow will help to determine both quantities in addition to the measurement of
the $\Pap/\Pp$ ratio at TeV energies in cosmic rays.
\vspace*{-0.3cm}\section{The Moon shadow effect}\label{sec:moon}
As already explained in section \ref{sec:intro}, the Moon shadow effect can be
exploited to measure the ratio of antimatter to matter in cosmic rays, using 
ground-based experiments with $\check\mathrm{C}$erenkov detectors, EAS arrays or muon track
detectors. 
The first two methods are sensitive to the total energy of the primary $E_{0}$,
so that $He^{4}$ and heavier nuclei participate significantly.
Muon experiments are sensitive to the primary energy per nucleon $E_{0}/A$,
which strongly favoures protons so that they measure mostly the $\Pap/\Pp$ 
ratio.\\
The L3+C detector acceptance ranges from zenith angle $\theta_{z}=0^{\circ}$ to 
almost $60^{\circ}$ and is given in Figure 3. The apparent 
position of the Moon is computed with the SLALIB~\cite{slalib} package, 
taking into account parallax corrections, at any time. The effect of the 
Earth's magnetic field has been simulated using the International Geomagnetic 
Reference Field (IGRF) model. Assuming a proton momentum, the
geomagnetic deflection $\Delta\overrightarrow{p}=\overrightarrow{p_{initial}}-
\overrightarrow{p_{detected}}$ is
computed. It is seen in Figure 3 that both the direction and the 
amount of deflection undergone by primary particles vary with the position of 
the Moon. It tends to blur the image of the Moon shadow. 
A new coordinate system, with axis parallel and orthogonal to the computed
deflexion has been defined to minimize this effect and will be referred to as
"deflection coordinate system".\\
\begin{center}
\begin{tabular}[!ht]{c c}
\includegraphics[height=6 cm,bb= 28 145 550 680]{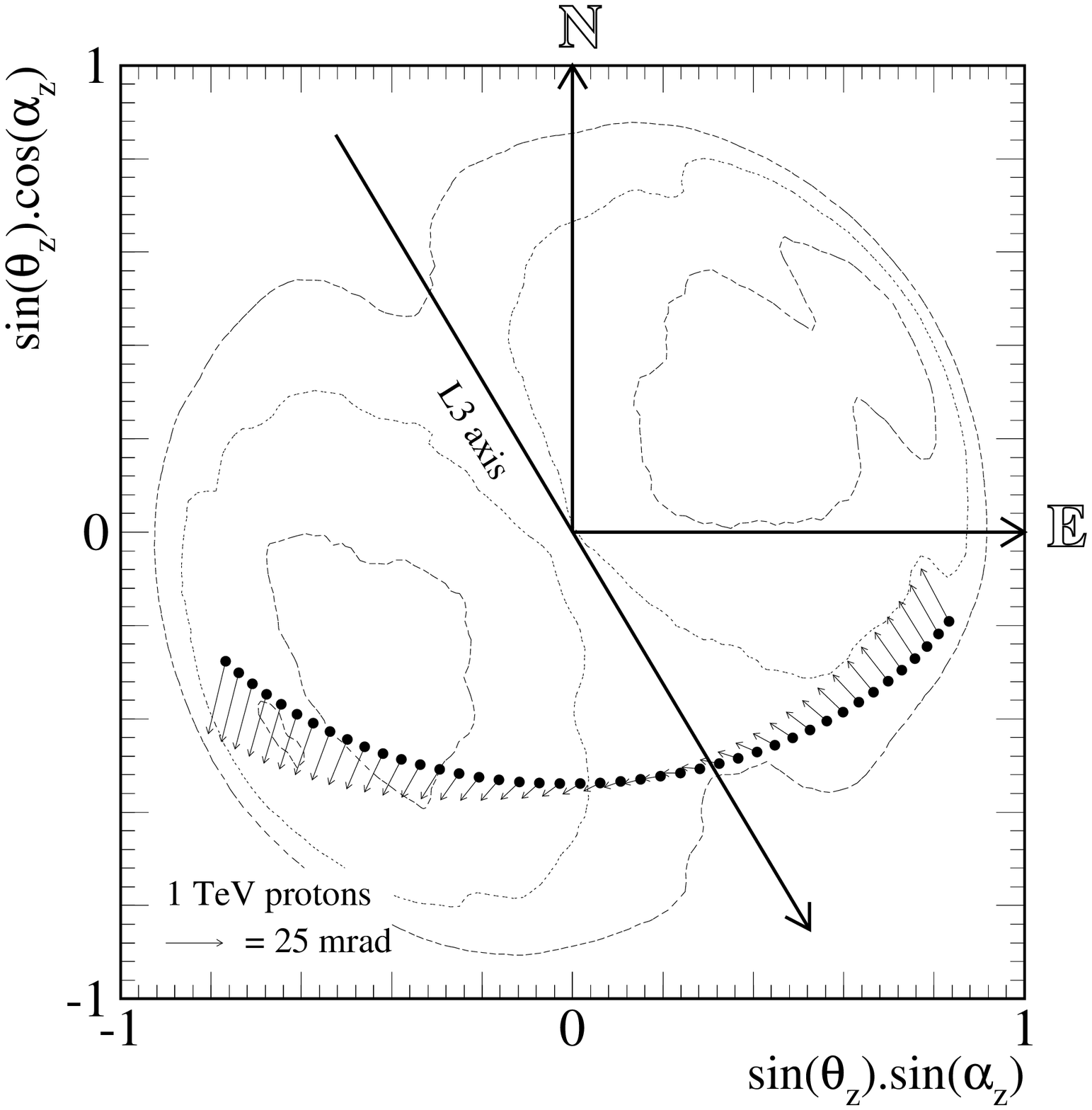}  &
\begin{minipage}{0.5\textwidth}
\vspace*{-6.cm}Figure 3 : \footnotesize{The L3+C acceptance. $\alpha_{z}$ and $\theta_{z}$ stand
respectively for the azimuth and zenith angles. Each pixel corresponds to a direction to 
the sky. A bipolar structure is observed due to 
the detector geometry. A transit of the Moon is indicated with dots. For each 
dot, the geomagnetic deflection is given by an arrow, whose length is 
proportional to the amount of deflection (see reference arrow in bottom-left 
corner).}
\end{minipage}\\
\end{tabular}
\end{center}
\addtocounter{figure}{1}
\section{Analysis procedure for the Moon shadow}\label{sec:anal}
\begin{figure}[!ht]
\hfill\begin{tabular}{c c c}
\includegraphics[width=4.7cm,height=4.4cm, bb= 27 150 565 647]{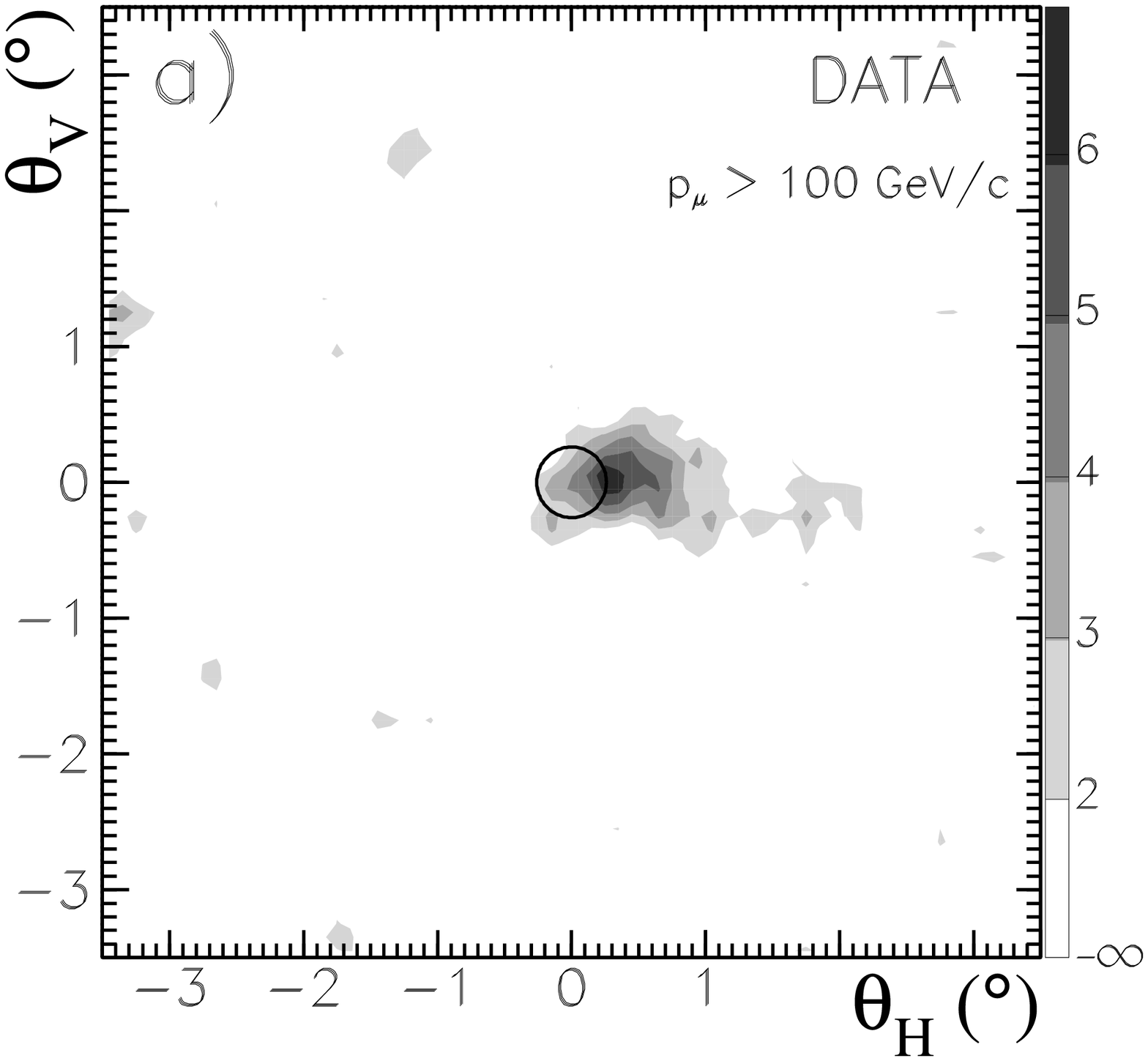} &
\includegraphics[width=4.7cm,height=4.4cm, bb= 27 150 565 647]{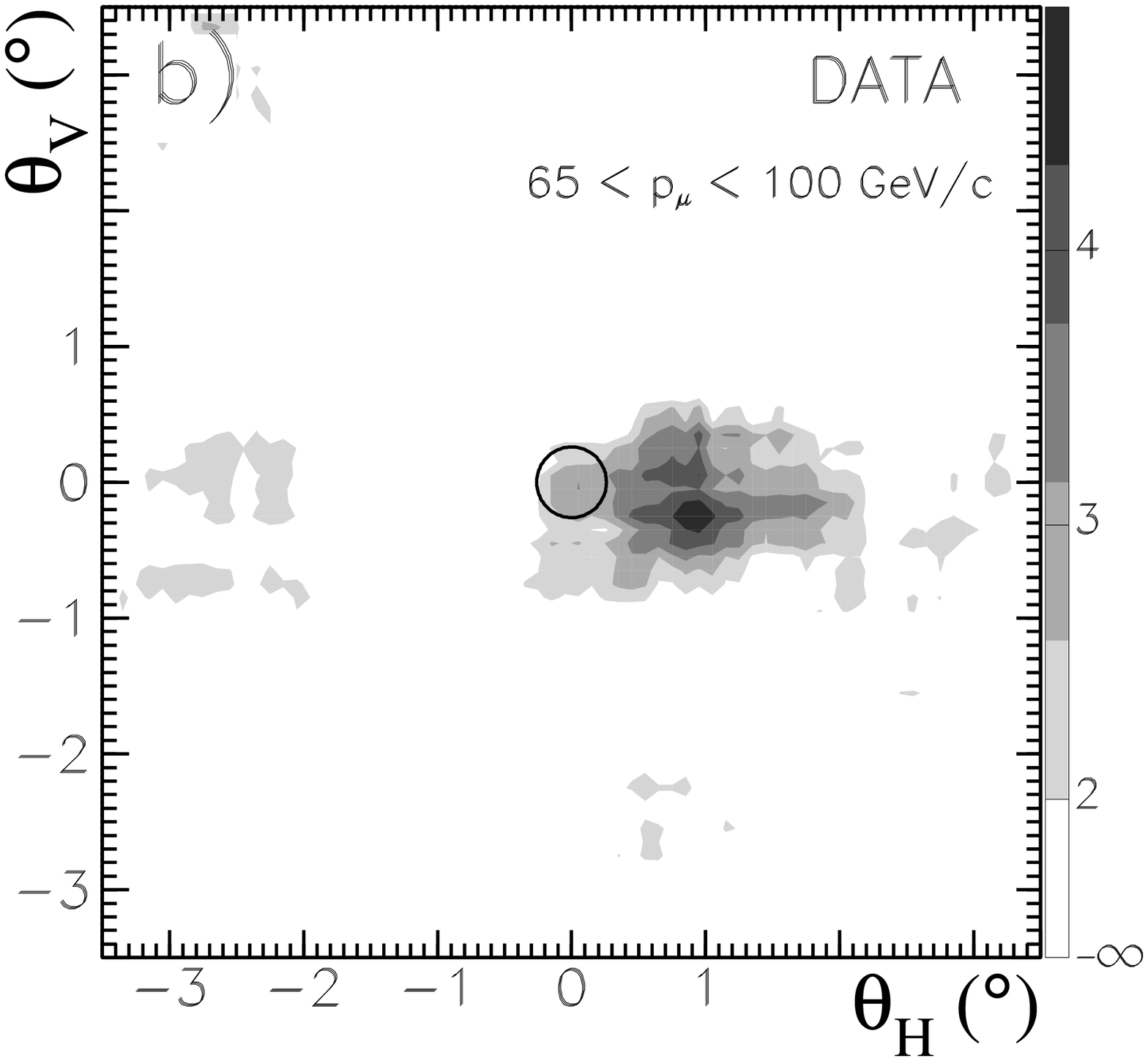} &
\includegraphics[width=4.7cm,height=4.4cm, bb= 16 150 565 647]{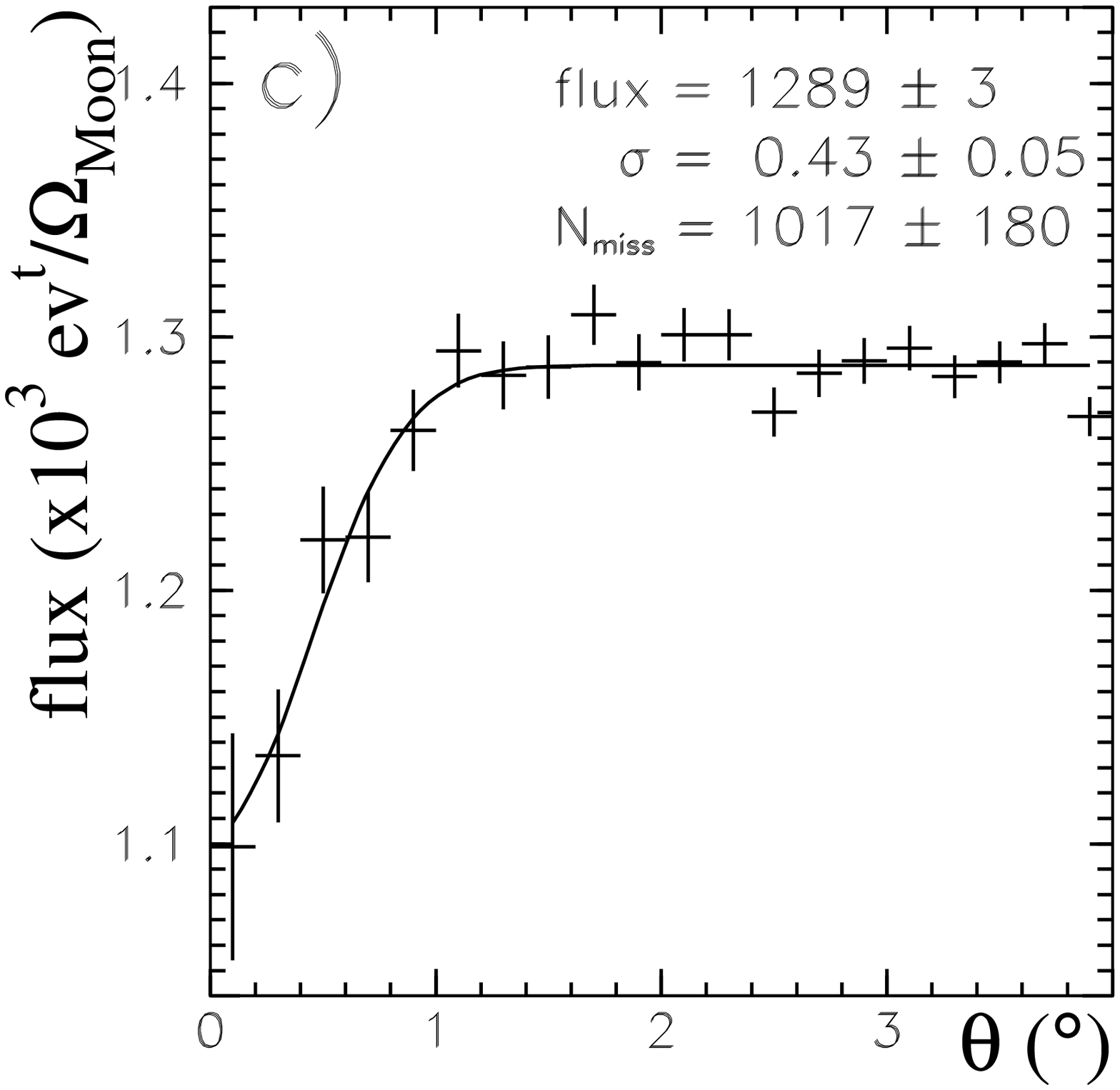}\\
\end{tabular}\hfill
\label{smooth}
\caption{\footnotesize{Moon shadow in the deflection coordinate system for 
HE (Figure a) and LE (Figure b) muons. The geomagnetic field effect is expected to act
along the horizontal axis. Color levels express the significance 
of the features, defined positive for a deficit. Figure (c) gives the flux versus the spatial angle 
$\theta$ from the centre of the deficit, with the Moon solid angle as solid
angle unit, for the whole muon sample (HE + LE).}}
\end{figure}

As a first step, the background has been subtracted from the data and a
smoothing algorithm applied. Results can be seen in figure \ref{smooth} for two
muon momentum ranges. In the deflection coordinate system, the Moon shadow is 
observed with a $7\sigma$ significance 
with high energy muons (HE: $p_{\mu}\,\geq 100\,GeV$, figure \ref{smooth}a) and $4.9\sigma$ significance
with low energy muons (LE: $65\,\leq\,p_{\mu}\,\leq 100\,GeV$, figure \ref{smooth}b). It is aligned along 
the horizontal axis as expected. Both the offset and the elongation due to the 
geomagnetic field are observed, with a more pronounced effect for low energy
muons as expected. Comparing the observed offset to the expected one (MC), the 
pointing error is deduced to be less than $0.1^{\circ}$. From Figure
\ref{smooth}c, an effective angular resolution of $0.43^{\circ}\pm 0.05$ is deduced for
the whole muon sample, including the geomagnetic field effect. A determination of
the effective angular resolution without the geomagnetic field effect is
performed by projecting both two-dimensional plots 
on the vertical axis, where the geomagnetic field effect is minimized. This
characteristic of the deflection coordinate system allows to extract angular
resolution values of $0.22^{\circ}\pm 0.05$ and $0.30^{\circ}\pm 0.07$
respectively for HE and LE muon samples.\\
A maximum likelihood fit has been performed on data in the deflection
coordinate system with all muons with momentum greater than 65 GeV (HE + LE), 
assuming that the data can be described as a planar background and 2 symmetric 
Gaussian deficits for protons and antiprotons:
\vspace*{-0.2cm}$$f(x,y)\,=\,\underbrace{ax+by+c}_{plane}
\,-\,\frac{N}{1+r}[\underbrace{F_{\Pp}(x_{0},y_{0},\sigma)}_{\Pp\ deficit}+
\underbrace{r}_{\Pap/\Pp\ ratio}\underbrace{F_{\Pap}(-x_{0}
,-y_{0},\sigma)}_{\Pap\ deficit}],$$
where parameters $x_{0}$ and $y_{0}$ express the offset due to the geomagnetic
field. Uncertainties on parameters have been computed with a Monte 
Carlo simulation. A deficit of $944\pm 320$ events is reported with $8\sigma$
significance. A preliminary $\Pap/\Pp$ ratio measurement is obtained: 
$r=-0.14\pm 0.15$. An upper limit has been set at 0.13 with $90\%$ C.L. using 
the "unified approach"~\cite{unif}.
\vspace*{-0.3cm}\section{Conclusion}
The L3+C experiment is not only one more experiment to observe a significant 
Moon shadow effect. Its good angular resolution has allowed for the first time 
to observe clearly the offset and the elongation of the Moon "muon shadow" 
expected from the geomagnetic field for 2 different muon energy ranges. 
Moreover, it allows a better separation of the proton and
antiproton deficits. A preliminary measurement of the $\Pap/\Pp$ ratio yields 
an upper limit of 0.13 at $90\%$ C.L. for the muon sample with $E_{\mu}\geq 65\,
GeV$.\\
This analysis is still preliminary since only one third of the data has been
reconstructed up to now. A complete reconstruction of the data is on the way. 
Improvements are also expected from a better description of the deficit and 
determination of the angular resolution from the study of dimuon events.

\end{document}